\documentclass{aastex6}

\usepackage{amsmath}
\usepackage{epstopdf}
\usepackage{textcomp}
\usepackage{graphicx}

\begin{document}


\title{The effect of gravitation on the polarization state of a light ray}


\newcommand{\myemail}{tanay_ghosh@hotmail.co.uk}
\author{Tanay Ghosh\altaffilmark{1} \and A. K. Sen\altaffilmark{1}}
\affil{Department of Physics\\
Assam University, Silchar-788011, Assam, India}

\begin{abstract}

In the present work detailed calculations have been carried out on the rotation of polarization vector of an electromagnetic wave due to the presence of gravitational field of a rotating body. This has been done using the general expression of Maxwell's equation in the
 curved space-time. Considering the far field approximation (i.e the impact parameter greater than the Schwarzschild
 radius and rotation parameter), the amount of rotation of polarization vector as a function of impact parameter has been
 obtained for a rotating body (considering Kerr geometry). Present work shows that, the rotation of polarization vector can
 not be observed in case of Schwarzschild geometry. This work also calculates the effect, considering prograde and retrograde
 orbit for the light ray. Although the present work demonstrates the effect of rotation of polarization vector but it confirms
 that, there would be no net polarization of electromagnetic wave due to the curved space-time geometry in Kerr field.

\end{abstract}



\section{\label{level1}Introduction\protect\\}
It was way back in 1958, when \cite{PhysRev.110.236} for the first time had  shown that the gravitational field of a
rotating body can actually rotate the direction of polarization vector of an electromagnetic wave passing in its field. The
author gave an idea about
how the gravitational field effects the orientation of the plane of polarization of the electromagnetic wave. The basic
concept behind the idea of \cite{PhysRev.110.236}, was to derive the four Maxwell's
equation \cite{jackson1962classical,born1999principles,stephani2004relativity,landau2013classical} and express the electric displacement vector
in terms of electric and magnetic field. He tried to figure out whether the gravitational field would have any effect on polarization or not?
It should be noted here, that the work by Balazs\cite{PhysRev.110.236} was carried out before the establishment of Kerr metric, so he described the
metric as:
\begin{align*}
 g_{\alpha \beta} & =(1-\frac{2\phi}{c^{2}})\delta_{\alpha \beta} & g_{00} & =(1+\frac{2\phi}{c^{2}}),\\
 \gamma & =-i(g_{i0})=(\frac{2G}{c^{3}R^{3}})L \times R
\end{align*}
where L represents the angular momentum of rotating mass and R represents the radius vector of rotating body. He investigated
 the problem further by considering the rotation of a
frame instead of a rotating body.\\
In the very next year another investigator \cite{PhysRev.118.1396}, had put forwarded his investigation on the propagation
 of electromagnetic wave
affected by gravitation. He tried to understand the similar physical phenomena of rotation of polarization vector due to the
gravitational effect in an isolated physical system or macroscopic medium. To resolve the physical problem, he took the geometrical optics approximation to estimate
the displacement vector and considered the case of very far field approximation. In that work from the change of the energy momentum tensor,
 he proved that energy conservation remains intact for the modified electric displacement vector. He described the electric displacement
 vector as:
 \begin{equation*}
  D_{i}=E_{i}+\varepsilon_{ik}E_{k}+\varepsilon_{ikl} g_{k} H_{l}
 \end{equation*}
Where $ D_{i}$ represents the displacement vector. He showed that when space becomes flat at very far distance, then $D_{i}\sim E_{i}$.
 This work also concentrated on the effect of the factors $\varepsilon_{ik}$ and $g_{k}$ on the rotation of polarization vector. The lowest order
approximation for $\varepsilon_{ik}$ and $g_{k}$ had been shown in terms of matter tensor (as described in \cite{PhysRev.118.1396})
 and concluded that the effect is very small. Finally,
the work gave the total rotation of polarization vector for a rotating frame and the author applied the result for some selected cases. This work clearly
established the rotation of the polarization vector during the deflection of light by gravitational field and also estimated the amount of rotation of
polarization vector.In the year 1977 \cite{pineault1977applications} estimated the rotation of the polarization vector with reference to a distance observer and calculated the image distortion due to the effect of the rotation of polarization vector. In 1980,  \cite{su1980effect} gave a comprehensive review of the work done till date, to calculate the amount of rotation which the polarization vector of a light ray will undergo when passing close to a massive rotating body({\it viz.} a black hole). In their review they discussed the cases, where some authors got non-vanishing results and others got vanishing results,  with the reasons for such differences.\\
In a  more recent work in last decade, \cite{sereno2004gravitational} investigated the rotation of polarization vector
due to gravitational field and gave the approximate result for gravitational Faraday rotation. He took the weak field approximation limit and
analytically presented the result for the gravitational Faraday rotation. The author claimed the model to be suitable for estimation of Faraday like rotation of polarization vector
up to third order correction to gravitational potential. He also discussed briefly the case for a gravitational lens moving with constant velocity.  Prior to that, a very systematic treatment on the image formation due to a moving gravitational lens was presented by \cite{paczynski1986gravitational}\\
As discussed above, it has been already well established that
 the direction of polarization vector gets rotated due to the effect of gravity \cite{carroll2004spacetime,PhysRev.110.236,vilenkin1984cosmic,fernandez2011x}.
 In general, space-time curvature created due to the presence of rotating massive
body, can cause the rotation of polarization vector of electromagnetic wave or light. The Sec.2 of this work, gives the general expression
for the rotation of the polarization vector for a linearly polarized light ray considering the far field approximation, which means the impact parameter is much greater than Schwarzschild radius and rotation parameter. The expression
for the electric displacement vector in the present work has been written with a knowledge of Kerr metric (\cite{kerr1963gravitational}), which was not possible by \cite{PhysRev.110.236} and \cite{PhysRev.118.1396} , as Kerr metric expression was published later. As discussed by \cite{sereno2004gravitational}
the model has significance in the micro-lensing phenomena, where the massive bodies act as moving micro-lenses in case of galactic gravitational lensing of light ray. However, this situation clearly differs from our case of interest, where we studied for a single ray the effect on the direction of polarization vector, by a rotating gravitational body. The authors \cite{su1980effect} developed their technique to calculate the rotation angle for two types of observers: in the first case the observer is in  a global inertial frame (GIF), ie in an asymptotic flat space and in the second case, the observer is a non-inertial one, at rest in a locally non-inertial frame(LNRF). The result obtained by them under the first case({\it viz.} GIF) can only be compared to the findings from our present study, as we limit our work to an observer in an asymptotic flat space only (i.e. at infinite distance from the massive rotating body). In Sec.5, the degree of polarization as may be expected due to this phenomena, has been obtained with a null result.
To derive the effect of gravitational field, here in the present work we derive the relation between  the electric displacement vector
and the electric field. A light source placed at asymptotically flat space emits electromagnetic wave, which travels through curved
space-time and is received by an observer at another asymptotically flat space. The effect of the gravitational field on the electromagnetic wave has
been calculated in the present work from the knowledge of the metric tensor representing the curved space-time. For simplicity the emitted
light from the source has been considered to be linearly (or plane) polarized for our convenience.

\section{\label{level2} General expression for the rotation of polarization vector of electromagnetic wave in curved space-time}

To derive the expression for desired rotation of polarization vector of electromagnetic wave, the expressions for the electromagnetic
displacement vector have to be calculated in generalized form, from the co-variant form of Maxwell equation. There expression can be written down as \cite{landau2013classical,stephani2004relativity,misner1973gravitation}:
\begin{subequations}
\begin{align}
\frac{\partial F_{ik}}{\partial x^{l}}+\frac{\partial F_{li}}{\partial x^{k}}+\frac{\partial F_{kl}}{\partial x^{i}} =0 \label{A1.1}\\
F^{ik}_{;k}  =\frac{1}{\sqrt{-g}}\frac{\partial}{\partial x^{k}}(\sqrt{-g}F^{ik}) \label{A1.2}
\end{align}
\end{subequations}

After the general expression of the displacement vector and magnetic induction have been established,
the general expression for rotation of polarization vector in curved space-time for a known gravitational field can be obtained.

\subsection{\label{sec:level2a}Expression for the electromagnetic displacement vector in curved space-time}
After writing the Maxwell equations in co-variant form, we may write down the expressions for electric field and displacement vector
in the corresponding three dimensional form as below \cite{landau2013classical,frolov2011spinoptics}:
 \begin{align}\label{Af}
E_{\alpha}&=F_{0 \alpha}           &   B_{\alpha \beta}&=F_{\alpha \beta}\\ \nonumber
D^{\alpha}&=-\sqrt{g_{00}}F^{0 \alpha}         &  H^{\alpha \beta}&=\sqrt{g_{00}}F^{\alpha \beta}
\end{align}
where E is electric component, B is magnetic component, D is the electric displacement vector and H is the magnetic induction of the electromagnetic
wave. The $\alpha$ and $\beta$ index run from 1 to 3 representing the spatial axis of the four dimensional space-time geometry. In this present work,
D and E are also represented in vectorial form in three dimensional geometry. Now from Eqn.(\ref{Af}) as outlined by
\cite{landau2013classical,frolov2011spinoptics} we may re-write, the following expression in a more detailed manner.
\begin{equation} \label{Ag}
\begin{split}
E_{\alpha} & =F_{0 \alpha}  \\
\Rightarrow E_{\alpha} & =g_{i0}g_{j\alpha}F^{ij}\\
\Rightarrow E_{\alpha} & =g_{i0}(g_{0\alpha}F^{i0}+g_{\alpha \beta}F^{i\beta})\\
\Rightarrow E_{\alpha} & =g_{00}(g_{0\alpha}F^{00}+g_{\alpha \beta}F^{0\beta})+g_{\gamma 0}(g_{0\alpha}F^{\gamma 0}+g_{\alpha \beta}F^{\gamma \beta})\\
\Rightarrow \frac{E_{\alpha}}{\sqrt{g_{00}}} & =\sqrt{g_{00}}g_{\alpha \beta}F^{0 \beta}+\frac{g_{0\alpha}g_{0\gamma}}{\sqrt{g_{00}}}F^{\gamma 0}+\frac{g_{\beta \alpha}g_{0\gamma}}{\sqrt{g_{00}}}F^{\gamma \beta}\\
\Rightarrow \frac{E_{\alpha}}{\sqrt{g_{00}}} & =-g_{\alpha \beta}(-\sqrt{g_{00}}F^{0\beta})+\frac{g_{0\alpha}g_{0\gamma}}{g_{00}}(-\sqrt{g_{00}}F^{\gamma 0})+\frac{g_{\beta \alpha}g_{0\gamma}}{g_{00}}H^{\gamma \beta}\\
\Rightarrow \frac{E_{\alpha}}{\sqrt{g_{00}}} & =-g_{\alpha \beta}D^{\beta}+\frac{g_{0\alpha}g_{0\gamma}}{g_{00}}D^{\gamma}+\frac{g_{\beta \alpha}g_{0\gamma}}{g_{00}}\gamma^{\mu \gamma}\gamma^{\nu \beta}H_{\mu \nu}\\
\Rightarrow \frac{E_{\alpha}}{\sqrt{g_{00}}} & =(\gamma_{\alpha \beta}-\frac{g_{0\alpha}g_{0\beta}}{g_{00}})D^{\beta}+\frac{g_{0\alpha}g_{0\gamma}}{g_{00}}D^{\gamma}+\frac{g_{0\gamma}}{g_{00}}\gamma^{\mu \nu}(-\partial^{\mu}_{\nu})H_{\mu \nu}\\
\Rightarrow \frac{E_{\alpha}}{\sqrt{g_{00}}} & =\gamma_{\alpha \beta}D^{\beta}-\frac{g_{0 \gamma}}{g_{00}}\gamma^{\mu \nu}H_{\mu \alpha}\\
\Rightarrow \frac{E_{\alpha}}{\sqrt{g_{00}}} & =D_{\alpha}-g^{\mu}H_{\alpha \mu}\\
\Rightarrow D_{\alpha} & =\frac{E_{\alpha}}{\sqrt{g_{00}}}+g^{\mu}H_{\alpha \mu}
\end{split}
\end{equation}
With the help of the identities of unit tensor \cite{landau2013classical,weinberg1972gravitation,frolov2011spinoptics} in three dimensional coordinate system, the Eqn.(\ref{Ag}) can be re-written in vector form as follows:
\begin{equation}
 D=\frac{E}{\sqrt{g_{00}}}+H \times g
 \label{Ah}
\end{equation}
So from Eqn.(\ref{Ah}), one can find out the components of the displacement vector for a given gravitational field and subsequently the
rotation of the polarization vector can be found out for a linearly polarized electromagnetic light ray.


\subsection{\label{level2b}Rotation of polarization vector of an electromagnetic wave in the  gravitational field of a rotating object}
\subsubsection{\label{level2b1}Light travelling along equatorial plane}
Let us assume that a linearly (or plane) polarized light ray is coming from an asymptotic flat space(represented by r=$-\infty$) and passing through
the gravitational field of a rotation object represented by the metric $g_{\mu\nu}$. After that the ray is received by an observer at r=$\infty$.
The displacement vector defined by Eqn.(\ref{Ah}) for the three dimensional coordinate system, in the present case has to be derived.
For our convenience, we define the geometry of the system as follows: The rotating gravitational mass is placed at the origin of three
dimensional cartesian co-ordinate system, with rotation axis defining the Z axis. The incoming light ray travelling along a plane
which is perpendicular to Z-axis. The direction of incoming light ray and the origin of co-ordinate
system together define the X-Y plane. The light ray approaches the origin with impact parameter \textquoteleft b\textquoteright. The original
direction of light ray marks the direction parallel to X axis (as illustrated by Fig.(\ref{Kerr})). The Y-axis is now defined as normal to X and
Z axis.\\
 In this geometry light ray is contained, within the equatorial plane of rotating body. The incoming light ray assumed to be plane polarized and it will
have its electric vector projected into the directions (Y,Z) with components $E_{y}$ and $E_{z}$. Thus the position angle of polarization ($\chi_{eq}$)
can be expressed by the relation:
\begin{equation}
 \chi_{eq}=\arctan (\frac{E_{y}}{E_{z}})
 \label{1}
\end{equation}

Now as the light propagates through the gravitational field, the effect of gravitation can be equated with the change in the refractive index of
the medium \cite{PhysRev.110.236,PhysRev.118.1396,sen2010more,Roy2014Sen}. As a result in the new material medium, the strength of electric field vectors will be now
estimated by the displacement vector $D_{y}$ and $D_{z}$. Thus the position angle of the polarization vector will be now estimated by the
formula,
\begin{equation}\label{v31}
 \tilde \chi_{eq}=\arctan (\frac{D_{y}}{D_{z}})
\end{equation}
 So the rotation of the polarization vector at each location of space, would be some over the total change from the position angle defined by Eqn.(\ref{1}).
Now Eqn.(\ref{v31}) only gives the position angle of the polarization vector at any point under the influence of gravitational field. To find out the total rotation of the polarization vector at any point, we have to calculate the rate of change of displacement vector at a particular point.
Now one can find the values of the components of the displacement vector along the Y and Z axes, with the help of Eqn.(\ref{Ah}) and Eqn.(\ref{1}).
We note that the electromagnetic wave is propagating parallel to the X axis and there will be no component of displacement vector or the magnetic
induction vector along the direction of propagation (which is X axis). So one can write:

\begin{subequations}
\begin{align}
    D_{y} & =\frac{E_{y}}{\sqrt{g_{00}}} +H_{z}g_{x} \label{EE1a}\\
    D_{z} & =\frac{E_{z}}{\sqrt{g_{00}}}-H_{y}g_{x} \label{EE1b}
\end{align}
\end{subequations}

Combining with Eqn.(\ref{v31}):
\begin{equation}\label{correqi}
\tilde \chi_{eq} =\arctan (\frac{D_{y}}{D_{z}})=\arctan \Big( \frac{\frac{E_{y}}{\sqrt{g_{00}}} +H_{z}g_{x}}{\frac{E_{z}}{\sqrt{g_{00}}}-H_{y}g_{x}}\Big)
\end{equation}
 where $\tilde \chi_{eq} $ is the position angle of polarization vector at a given point, under the influence of gravitational field. To find the change of electric displacement vector, we must know, the components of electric vector are $E_{y}$ and $E_{z}$ along the Y and Z axes at that at point $r=-\infty$.
 Now, if we assume that the polarization vector makes an angle $\chi_{eq}  $  with Y axis at the point $r=-\infty$, then the components of electric vector will be related by the equation $E_{y}=\xi_{eq} E_{z}$ (say), $\xi_{eq}$ is given by,
\begin{equation}\label{2baeq1}
 \chi_{eq} =\arctan (\frac{E_{y}}{E_{z}})=\arctan(\xi_{eq})
\end{equation}
Now the different components of magnetic field (H)  at the starting point of incoming ray can be written in terms of the electric field by the following relation
 $\vec H_{y}=\hat{n}\times \vec E_{y}$ and $\vec H_{z}=\hat{n}\times \vec E_{z}$, (page-112 of \cite{landau2013classical}), where $\hat{n}$
 is the normal unit vector along the direction of propagation(in this case parallel to X axis). So taking only the magnitude of magnetic field from Eqn.(\ref{EE1a}) and Eqn.(\ref{EE1b}) and taking note of our earlier assumption that $E_{y}=\xi E_{z}$, we may write,

\begin{subequations}
\begin{align}
 D_{y} & =E_z\{\frac{\xi_{eq}}{\sqrt{g_{00}}}+ g_{x}\}\label{correq1a}\\
 D_{z} & =E_z\{\frac{1}{\sqrt{g_{00}}}- \xi_{eq} g_{x}\}\label{correq1b}
\end{align}
\end{subequations}

As a result we may finally write,

\begin{equation}\label{correq2}
\begin{split}
 \tilde \chi_{eq} &=\arctan (\frac{D_{y}}{D_{z}}) \\
 &=\arctan (\frac{\xi_{eq} +\sqrt{g_{00}}g_{x}}{1-\xi_{eq} \sqrt{g_{00}}g_{x}})
\end{split}
\end{equation}
 Eqn.(\ref{correq2}) gives the position angle of the rotated polarization vector of the electromagnetic wave.
 Now at initial point ($r=-\infty$ in our case) $D_{y}=E_{y}$ and $D_{z}=E_{z}$. The general equation for $D_{y}$, $D_{z}$ at any point r(say) it will be,
 \begin{subequations}
\begin{align}
 D_{y} & =\frac{\xi_{eq} E_z}{\sqrt{g_{00}(r)}}+ E_{z} g_{x}(r)\label{Eqn2bn1}\\
 D_{z} & =\frac{E_z}{\sqrt{g_{00}(r)}}- E_{z}\xi_{eq} g_{x}(r)\label{Eqn2bn2}
\end{align}
\end{subequations}
 From Eqn.(\ref{Eqn2bn1}) and Eqn.(\ref{Eqn2bn2}) it is clear that both $D_{y}$ and $D_{z}$ are function of position only.
\subsubsection{\label{level2bb}Light propagation along rotation axis}
 If the light ray propagates along the rotation axis i.e. along the defined Z axis, then the will have its electric vector projected into the directions(X,Y) with components $E_{x}$ and $E_{y}$. Thus the position angle
of polarization $(\chi_{ax})$ can be expressed as:
\begin{equation}\label{1pp}
\chi_{ax}=\arctan (\frac{E_{x}}{E_{y}})
\end{equation}
Now after the light ray propagates out through the gravitational field, the strength of electric field vectors will be estimated by the displacement
vectors \cite{PhysRev.110.236,PhysRev.118.1396,sen2010more,Roy2014Sen} $D_{x}$ and $D_{y}$. Thus the position angle of polarization
vector will be estimated by the formula,

\begin{equation}\label{v31pp}
 \tilde \chi_{ax}=\arctan (\frac{D_{x}}{D_{y}})
\end{equation}

As it has been shown in previous Sec.\ref{level2b1} one can find the values of $D_{x}$ and $D_{y}$,with the assumption that
$E_{x}=\xi_{ax} E{y}$ at any point r(say), by the following relations

 \begin{subequations}
\begin{align}
 D_{x} & =\frac{\xi_{ax} E_y}{\sqrt{g_{00}(r)}}+ E_{y} g_{z}(r)\label{Eqn2bn1pp}\\
 D_{y} & =\frac{E_y}{\sqrt{g_{00}(r)}}- E_{y}\xi_{ax} g_{z}(r)\label{Eqn2bn2pp}
\end{align}
\end{subequations}
Here again from Eqn.(\ref{Eqn2bn1pp}) and Eqn.(\ref{Eqn2bn2pp}), it is clear that both $D_{x}$ and $D_{y}$ are function of
position ($r$) only.
\section{\label{level3}Metric element for light ray travelling along and perpendicular to equatorial plane}

 In this section the effect of gravitational field of a rotating object on the total amount of rotation of polarization vector will be discussed.
 In our case the electric displacement vector as in Eqn.(\ref{correq1a}) and Eqn.(\ref{correq1b}) (and Eqn.\ref{Eqn2bn1pp} and \ref{Eqn2bn2pp})
 have been defined in cartesian co-ordinate system, so the situation demands that we must express the metric $¨g¨$ or the corresponding
line element in cartesian form.  The authors \cite{su1980effect} had suitably parameterized the Kerr line element, to look for symmetry and to find a solution to the problem. They considered two geometries, the light ray passing parallel to the rotation axis and ray passing along the equatorial plane. The Kerr field line element is generally expressed in spherical co-ordinates as given bellow \cite{kerr1963gravitational,visser2007kerr},
\begin{widetext}
\begin{multline}\label{3ceq1}
ds^{2}=\frac{\Sigma}{\Delta}dr^{2}-\frac{\Delta}{\Sigma}[dt-a\sin ^{2}\theta d\phi]^{2}+\frac{\sin ^{2}\theta}{\Sigma}[(r^{2}+a^{2})d\phi-adt]^{2}+\Sigma d\theta ^{2}
\end{multline}
 \end{widetext}
 where $\Delta =r^{2}-r_{g}r+a^{2}$ and $\Sigma= r^{2}+a^{2}\cos ^{2}\theta$. Bellow we shall consider the above two cases of geometry and evaluate the ´metric´ in cartesian form for the two cases separately.
\subsection{\label{level3a}Light ray travelling along equatorial plane}
In this case we have taken the light ray to be contained in the equatorial plane of the system, so $\sin \theta =1$ and $\cos \theta =0$.
 Taking this into account, we have $\Sigma = r^{2}$.
Thus the above Eqn.(\ref{3ceq1}) can be rewritten as:
\begin{equation}\label{3ceq2}
 \begin{split}
 ds^{2}&=-(1-\frac{r_{g}}{\sqrt{x^2+y^2}})dt^{2}\\
 &+\big(\frac{x^{2}}{r^{2}-r_{g}r+a^{2}}+\frac{y^{2}(2a^{2}+r_{g}r+r^{4}-r^{2})}{r^{6}}\big)dx^{2}\\
 &+\big(\frac{y^{2}}{r^{2}-r_{g}r+a^{2}}+\frac{x^{2}(2a^{2}+r_{g}r+r^{4}-r^{2})}{r^{6}}\big)dy^{2}\\
 &+2\frac{r_{g}ay}{(x^2+y^2)^{\frac{3}{2}}}dtdx-2\frac{r_{g}ax}{(x^2+y^2)^{\frac{3}{2}}}dtdy\\
 &+2xy(\frac{1}{r^{2}-r_{g}r+a^{2}}-\frac{2a^{2}+r_{g}r+r^{4}-r^{2}}{r^{6}})dxdy
\end{split}
\end{equation}

\begin{figure*}[!htb]
\centerline{\includegraphics[width=10cm]{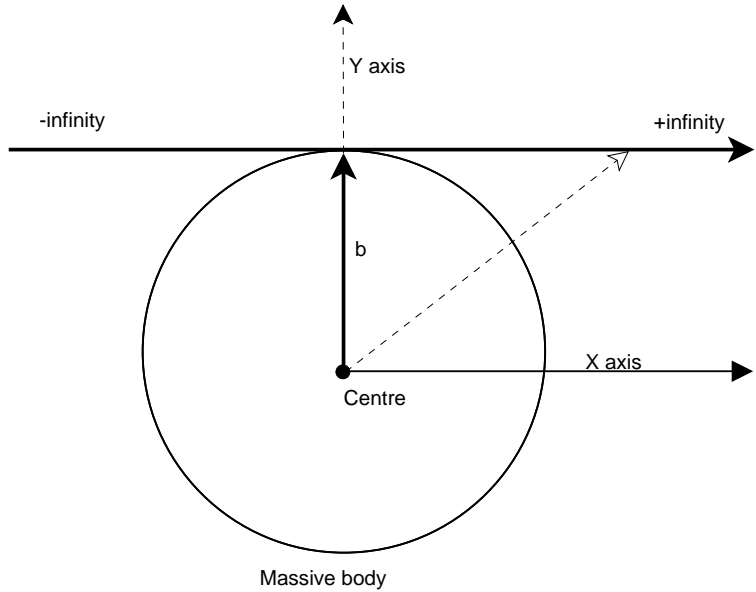}}
\caption{Shows schematic diagram of an electromagnetic wave passing close to a gravitational mass with impact factor `b'
\label{Kerr}}
\end{figure*}

So the metric for the Kerr line element will be,
\begin{eqnarray}\label{3ceq3}
g_{\mu \nu}=\left(
\begin{array}{cccc}
-(1-\frac{r_{g}}{\sqrt{x^2+y^2}}) & \frac{r_{g}ay}{(x^2+y^2)^{\frac{3}{2}}} & -\frac{r_{g}ax}{(x^2+y^2)^{\frac{3}{2}}} & 0\\
\frac{r_{g}ay}{(x^2+y^2)^{\frac{3}{2}}} & \big(\frac{x^{2}}{r^{2}-r_{g}r+a^{2}}+\frac{y^{2}(2a^{2}+r_{g}r+r^{4}-r^{2})}{r^{6}}\big) & xy(\frac{1}{r^{2}-r_{g}r+a^{2}}-\frac{2a^{2}+r_{g}r+r^{4}-r^{2}}{r^{6}}) & 0\\
-\frac{r_{g}ax}{(x^2+y^2)^{\frac{3}{2}}} & xy(\frac{1}{r^{2}-r_{g}r+a^{2}}-\frac{2a^{2}+r_{g}r+r^{4}-r^{2}}{r^{6}}) & \big(\frac{y^{2}}{r^{2}-r_{g}r+a^{2}}+\frac{x^{2}(2a^{2}+r_{g}r+r^{4}-r^{2})}{r^{6}}\big) & 0\\
0 & 0 & 0 & 0
\end{array}\right)
\end{eqnarray}
Here from Eqn.( \ref{3ceq3}) the $g_{00}$ and $g_{x}$ (or $g_{0x}$) can be easily identified as:
\begin{align}\label{3ceq4}
 g_{00}&=(1-\frac{r_{g}}{\sqrt{x^2+y^2}})\\
 g_{x}& =\frac{r_{g}ay}{(x^{2}+y^{2})^{\frac{3}{2}}}
\end{align}
 The spatial component $\gamma_{\alpha \beta}$,( \cite{landau2013classical}) could be extracted from the Eqn.(\ref{3ceq2}) and written as:
 \begin{eqnarray}\label{3ceq4m}
\gamma_{\alpha \beta}=\left(
\begin{array}{cccc}
\big(\frac{x^{2}}{r^{2}-r_{g}r+a^{2}}+\frac{y^{2}(2a^{2}+r_{g}r+r^{4}-r^{2})}{r^{6}}\big) & xy(\frac{1}{r^{2}-r_{g}r+a^{2}}-\frac{2a^{2}+r_{g}r+r^{4}-r^{2}}{r^{6}}) & 0\\
xy(\frac{1}{r^{2}-r_{g}r+a^{2}}-\frac{2a^{2}+r_{g}r+r^{4}-r^{2}}{r^{6}}) & \big(\frac{y^{2}}{r^{2}-r_{g}r+a^{2}}+\frac{x^{2}(2a^{2}+r_{g}r+r^{4}-r^{2})}{r^{6}}\big) & 0\\
0 & 0 & 0 \end{array}\right)
\end{eqnarray}
 For the geometry as explained in Fig.(\ref{Kerr}) for impact parameter $b$, we can write (approximately):
\begin{equation}\label{apporoximation}
 r^{2}=x^{2}+b^{2}
\end{equation}
\medskip
As the amount of deflection is small, we have used y=b for entire trajectory of light. We also note $r\gg b$.
Applying the result for sufficiently large $r$ in Eqn.(\ref{Eqn2bn1}) and Eqn.(\ref{Eqn2bn2}), we can say that displacement
vector(D) is a function of x only. So the displacement vector at point (x+$\bigtriangleup$x) will be,

 \begin{subequations}
\begin{align}
 D_{y} & =E_{z}\Bigg( \frac{\xi}{\sqrt{1-\frac{r_{g}}{\sqrt{(x+\bigtriangleup x)^{2}+b^{2}}}}}+ \frac{r_{g}ab}{(\sqrt{(x+\bigtriangleup x)^2+b^2})^3}\Bigg)\label{Eqn3cn1}\\
 D_{z} & =E_{z}\Bigg( \frac{1}{\sqrt{1-\frac{r_{g}}{\sqrt{(x+\bigtriangleup x)^{2}+b^{2}}}}}+ \frac{\xi r_{g}ab}{(\sqrt{(x+\bigtriangleup x)^2+b^2})^3}\Bigg)\label{Eqn3cn2}
\end{align}
\end{subequations}
Thus the change of $D_{y}$ is $D_{y}(x+\bigtriangleup x)-D_{y}(x)$ for the change in position co-ordinate by $\bigtriangleup x$. Therefore, the rate of change would be given as:
\begin{widetext}
\begin{align}\label{Eqn3cn3}
\lim_{\bigtriangleup x\rightarrow 0}\frac{D_{y}(x+\bigtriangleup x)-D_{y}(x)}{\bigtriangleup x}=E_{z}\lim_{\bigtriangleup x\rightarrow 0}\frac{( \frac{\xi}{\sqrt{1-\frac{r_{g}}{\sqrt{(x+\bigtriangleup x)^{2}+b^{2}}}}}+ \frac{r_{g}ab}{(\sqrt{(x+\bigtriangleup x)^2+b^2})^3})-( \frac{\xi}{\sqrt{1-\frac{r_{g}}{\sqrt{(x)^{2}+b^{2}}}}}+ \frac{r_{g}ab}{(\sqrt{(x)^2+b^2})^3})}{\bigtriangleup x}
\end{align}
\end{widetext}

We note here $E_{y}$ and $E_{z}$ are independent of $x$. Now after some steps of simplification, the rate of change of electric
displacement vector along the Y axis can be calculated as:
\begin{widetext}
\begin{equation}\label{Eqn3cn4}
\frac{dD_{y}}{dx}=E_{z}\Bigg[\frac{\xi}{2}\frac{x}{\sqrt{\sqrt{x^2+b^2}}\sqrt{\sqrt{x^2+b^2}-r_{g}}}\Big( \frac{1}{\sqrt{x^2+b^2}}-\frac{1}{\sqrt{x^2+b^2}-r_{g}}\Big)-3r_{g}ab\frac{x}{(x^2+b^2)^{\frac{5}{2}}} \Bigg]
\end{equation}
\end{widetext}
 It is easy to observe that the function $\frac{dD_{y}}{dx}$ is an even function so the total change of electric
displacement vector from initial point ($r=-\infty$) to final point ($r=\infty$) will be,
\begin{widetext}
\begin{equation}\label{Eqn3cn5}
\begin{split}
\bigtriangleup D_{y}|^{\infty}_{x=-\infty}  &=2\int_{0}^{\infty} E_{z}\Bigg[\frac{\xi}{2}\frac{x}{\sqrt{\sqrt{x^2+b^2}}\sqrt{\sqrt{x^2+b^2}-r_{g}}}\Big( \frac{1}{\sqrt{x^2+b^2}}-\frac{1}{\sqrt{x^2+b^2}-r_{g}}\Big)-3r_{g}ab\frac{x}{(x^2+b^2)^{\frac{5}{2}}} \Bigg]dx\\
 &=2\int_{0}^{\infty} E_{z}\xi \Bigg[ \frac{r_{g}}{2}\frac{x}{(x^{2}+b^{2})^{\frac{3}{4}}(\sqrt{x^{2}+b^{2}}-r_{g})^{\frac{3}{2}}}\Bigg]dx\\
 &=2E_{z}\xi \bigg[ -\frac{\sqrt{\sqrt{x^2+b^2}}}{\sqrt{\sqrt{x^2+b^2}-r_{g}}}+\frac{r_{g}ab}{(x^2+b^2)^{\frac{3}{2}}}\bigg]^{\infty}_{0}\\
 &=2E_{z}\xi \bigg[-1+\frac{\sqrt{b}}{\sqrt{b-r_{g}}}-\frac{r_{g}a}{b^2}\bigg]
\end{split}
\end{equation}
\end{widetext}
From Eqn.(\ref{Eqn3cn5}), we can get the electric vector at the final point ($r=+\infty$), by adding the total change
 ($\bigtriangleup D_{y}|^{\infty}_{x=-\infty}$) to the initial value of electric vector and it is given as:
\begin{equation}\label{Eqn3cn6}
\begin{split}
D_{y}(x=+\infty)&=E_{y}+\bigtriangleup D_{y}\big|^{\infty}_{x=-\infty} \\
&=E_{z}\xi \Big[1+ 2\big[-1+\frac{\sqrt{b}}{\sqrt{b-r_{g}}}-\frac{r_{g}a}{b^2}\big]\Big]
\end{split}
\end{equation}
Similarly from Eqn.(\ref{Eqn3cn2}), we can find the expression for electric vector along Z axis and it can be expressed as:
 \begin{equation}\label{Eqn3cn7}
 \begin{split}
D_{z}(x=+\infty)&=E_{z}+\bigtriangleup D_{z}\big|^{\infty}_{x=-\infty}\\
&=E_{z}\Big[1+ 2\big[-1+\frac{\sqrt{b}}{\sqrt{b-r_{g}}}+\frac{r_{g}a}{b^2}\big]\Big]
\end{split}
\end{equation}
It is also clear from the definition of the problem that $D_{y}(x=\infty)$ and $D_{z}(x=\infty)$ are nothing but the
components of electric field as will be recognized at asymptotically flat space $(x=\infty)$. So the total rotation of
the polarization vector for the light ray travelling along the equatorial plane would be(combining
Eqn.(\ref{2baeq1},\ref{correq2},\ref{Eqn3cn6},\ref{Eqn3cn7})),
\begin{equation}\label{Eqn3cn8}
\begin{split}
\Delta\chi_{eq} & = \arctan( \xi )-\arctan(\frac{D_{y}(at\ r=\infty)}{D_{z}(at\ r=\infty)})\\
     & = \arctan(\xi )- \arctan \xi \frac{\big[1+ 2[-1+\frac{\sqrt{b}}{\sqrt{b-r_{g}}}-\frac{r_{g}a}{b^2}]\big]}{\big[1+2[-1+\frac{\sqrt{b}}{\sqrt{b-r_{g}}}+\frac{r_{g}a}{b^2}]\big]}\\
     & = \arctan(\xi )- \arctan \xi\frac{1-\frac{2\frac{r_{g}a}{b^2}}{2\frac{\sqrt{b}}{\sqrt{b-r_{g}}}-1}}{1+\frac{2\frac{r_{g}a}{b^2}}{2\frac{\sqrt{b}}{\sqrt{b-r_{g}}}-1}}\\
     & = \arctan \frac{2\xi \frac{2\frac{r_{g}a}{b^2}}{2\frac{\sqrt{b}}{\sqrt{b-r_{g}}}-1}}{(1+\frac{2\frac{r_{g}a}{b^2}}{2\frac{\sqrt{b}}{\sqrt{b-r_{g}}}-1})+\xi^{2}(1-\frac{2\frac{r_{g}a}{b^2}}{2\frac{\sqrt{b}}{\sqrt{b-r_{g}}}-1})}\\
     &= \arctan \frac{2\tan(\chi_{eq} ) \frac{2\frac{r_{g}a}{b^2}}{2\frac{\sqrt{b}}{\sqrt{b-r_{g}}}-1}}{(1+\frac{2\frac{r_{g}a}{b^2}}{2\frac{\sqrt{b}}{\sqrt{b-r_{g}}}-1})+\tan^{2}(\chi_{eq})(1-\frac{2\frac{r_{g}a}{b^2}}{2\frac{\sqrt{b}}{\sqrt{b-r_{g}}}-1})}
\end{split}
\end{equation}
The Eqn.(\ref{Eqn3cn8}) gives the general expression for rotation of polarization vector of a light ray travelling along the
 equatorial plane of a Kerr mass, where $\chi_{eq} $ is the initial value of the position angle of polarization (electric vector) when light leaves its source at r= -$\infty$. We can clearly see when we substitute $a=0$
(representing Schwarzschild field), the value of $\Delta \chi_{eq}$ becomes zero from Eqn.(\ref{Eqn3cn8}). This confirms that, for
Schwarzschild geometry there will be no rotation of polarization vector.
\subsection{\label{level3b}Light ray travelling parallel to the rotation axis}
 Now, we consider that the light ray propagates parallel to the rotation axis i.e. the Z axis according to our definition.
In this case we could consider that the light ray is in the Y-Z plane and the choice of the X axis becomes arbitrary. Now looking back
 to Boyer-Lindquist equation;
Eqn.(\ref{3ceq1}) one would have $d\phi=0$ and the Eqn.(\ref{3ceq1}) will be reduced to:
\begin{multline}\label{3ceq1pp}
ds^{2}=\frac{\Sigma}{\Delta}dr^{2}-\frac{\Delta}{\Sigma}dt^{2}+\frac{\sin ^{2}\theta}{\Sigma}(adt)^{2}+\Sigma d\theta ^{2}
\end{multline}
where $\Sigma$ and $\Delta$ are previously defined in Eqn.(\ref{3ceq1}). Following the same steps as shown in
Sec.\ref{level3a} one could simplify
Eqn.(\ref{3ceq1pp}) and it is given by:
\begin{equation}\label{3ceq2pp}
 \begin{split}
 ds^{2}&=-(\frac{\Delta}{\Sigma}+\frac{y^{2}a^{2}}{r^{2}\Sigma{2}})dt^{2}\\
 &+(\frac{\Sigma y^2}{\delta r}+\frac{\Sigma}{z^{2}}(1-\frac{y}{r})^{2})dy^{2}\\
 &+(\frac{\Sigma z^{2}}{\Delta r}-(\frac{y}{r})^{2})dz^{2}\\
 &+2(\frac{\Sigma y^{2}}{\Delta r}+\frac{\Sigma y}{zr}(1-\frac{y}{r}))dydz
\end{split}
\end{equation}
Now following the same assumption as in Sec.\ref{level3a}, that for the large value of r$(\gg b$, impact parameter) we can
write:
\begin{equation*}
 r^{2}=z^{2}+b^{2}
\end{equation*}
 But from Eqn.(\ref{3ceq2pp}) we see that here $g_{z0}\equiv g_{z}=0$, so from Eqn.(\ref{v31pp}) with the help of
Eqn.(\ref{Eqn2bn1pp}) and Eqn.(\ref{Eqn2bn2pp}) we can write that
\begin{equation}
 \tilde \chi_{ax}=\arctan (\frac{E_{x}}{E_{y}})=\chi_{ax}
\end{equation}
So there will be no change of position of polarization angle ($\chi$), when electromagnetic radiation propagates parallel to
the rotational axis of a rotating mass.\\
 In the work authored by \cite{PhysRev.110.236} it had been shown that the rotation of  of polarization vector
is non-zero when light ray travels parallel to the rotation axis in Kerr field. Thus, our result here is definitely
contradicting and totally opposite to what \cite{PhysRev.110.236}reported in 1958. However \cite{PhysRev.110.236} did not
calculate the value of rotation of polarization vector when light ray is travelling along the equatorial plane.
\section{\label{pulsar} Some sample calculations for the rotation of polarization vector; when light ray travels along equatorial plane}
\begin{figure*}[!htb]
\centerline{\includegraphics[width=14cm]{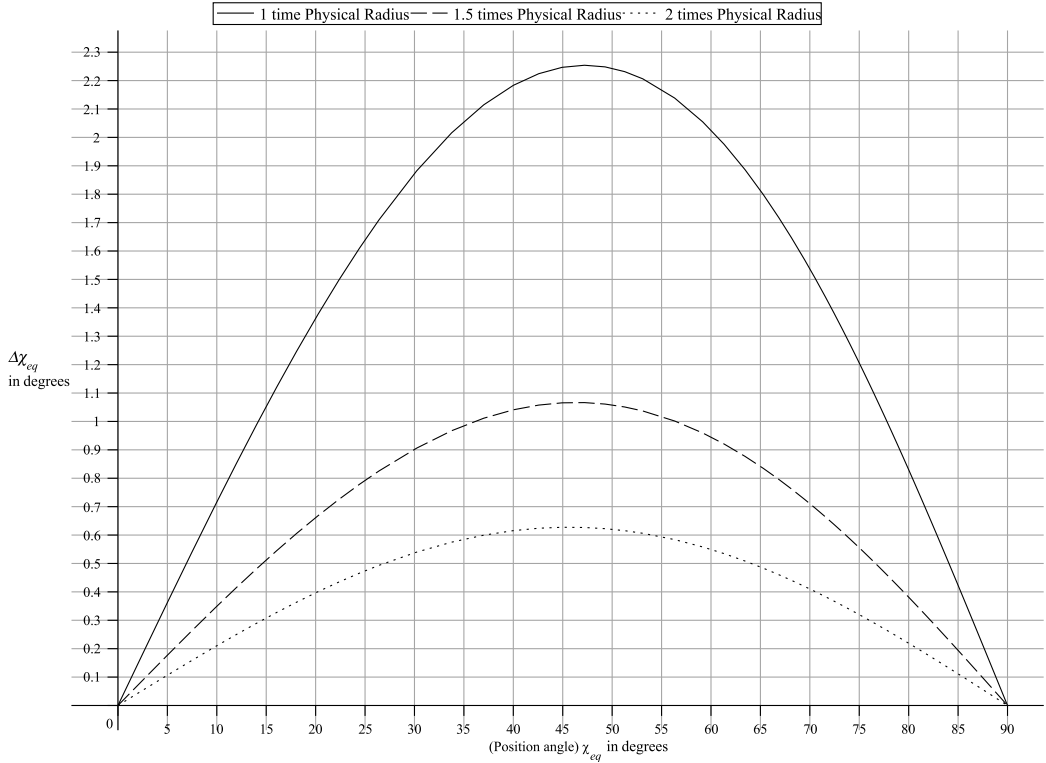}}
\caption{Shows a plot of $\Delta \chi_{eq}$ against Position angle ($\chi_{eq}=\arctan(\frac{E_{y}}{E_{z}})$) where $\chi_{eq}$ represents the orientation of initial polarization vector at $r=-\infty$. The $\Delta \chi_{eq}$ is the rotation in position angle. The three curves represent three plots with impact parameter equal to 1, 1.5, and 2  times the Physical Radius. 
 \label{Impact}}
\end{figure*}
In order to study the variation  of $\Delta \chi_{eq}$ with respect to $\chi_{eq}$, we take its  first order derivative from Eqn.(\ref{Eqn3cn8}).
Thus we see $\Delta \chi_{eq}$ becomes extremum when, $\frac{d \Delta \chi_{eq}}{d \chi_{eq}}=0$. Thus we have,
\begin{widetext}
\begin{equation}\label{extremum}
 \frac{d \Delta \chi _{eq}}{d \chi_{eq}}=\sec ^{2} \chi \Bigg [ \frac{4r_{g}}{b^{2}(\frac{2\sqrt{b}}{\sqrt{b-r_{g}}})} a (1+ \frac{4r_{g}}{b^{2}(\frac{2\sqrt{b}}{\sqrt{b-r_{g}}})} a) -\frac{4r_{g}}{b^{2}(\frac{2\sqrt{b}}{\sqrt{b-r_{g}}})} a \tan^{2}(\chi_{eq}) (1- \frac{4r_{g}}{b^{2}(\frac{2\sqrt{b}}{\sqrt{b-r_{g}}})} a) \Bigg]=0
\end{equation}
\end{widetext}
From Eqn.(\ref{extremum}) we can notice that $\sec \chi_{eq}$ can not be zero and solving the other part of equation to zero we have,
\begin{equation}\label{extreamumf}
 \chi_{eq} =\sqrt{\frac{1-\frac{4r_{g}}{b^{2}(\frac{2\sqrt{b}}{\sqrt{b-r_{g}}})} a}{1+ \frac{4r_{g}}{b^{2}(\frac{2\sqrt{b}}{\sqrt{b-r_{g}}})}a}}
\end{equation}
 From Eqn.(\ref{extreamumf}) we can find out the value of $ \chi_{eq}$ for which the rotation in position angle will be extremum.  Taking a second derivative we can show that, this value represented by Eqn.(\ref{extreamumf}), gives the condition for maximum. 

\subsubsection{Case I: $E_{y}=0$ i.e. $\xi=0$ or $\chi_{eq} $=0}
 If we consider $E_{y}=0$ then $\xi$ will be zero. Now from Eqn.(\ref{Eqn3cn8}) the total rotation of polarization vector will be
zero.
\subsubsection{Case II: $E_{y}=E_{z}$ i.e. $\xi =1$ or $\chi_{eq} $=45\textdegree}
Considering the both components are equal the initial position angle of polarization vector will be 45\textdegree. In this case
from Eqn.(\ref{Eqn3cn8}) we could write:
\begin{equation}\label{Eqn3cn845}
\Delta \chi_{eq} 
=\arctan \frac{2\frac{r_{g}a}{b^2}}{2\frac{\sqrt{b}}{\sqrt{b-r_{g}}}-1}
\end{equation}
Sun is rotating with an angular speed $1.664\times10^{-4}$ \textdegree /s  (degrees per second) and it has moment of inertia
$5.8\times 10^{46}$ kg. $m^{2}$. Accordingly, we can calculate the rotational parameter `a' for Sun as 1.69749 km. Considering
the impact parameter `b' of Sun as the physical radius of Sun (0.7$\times 10^{6}$km.), from Eqn.(\ref{Eqn3cn845})we can
estimate the value of $\Delta \chi_{eq}$ for Sun as 4.2873 micro-arcsec. One can say, there will be no observable change for the rotation
 of polarization vector in case of Sun. However, when we apply our  Eqn.( \ref{Eqn3cn845}) to other fast rotating compact
objects, we find  the result is quite fascinating. For these compact rotating objects, the light ray passes closer to the
centre of mass and stronger field is experienced by light ray. So we obtain much higher values for the total rotation
of the polarization vector. We take couple of test cases and let us choose the cases of two millisecond
pulsars PSR J 1748-2446ad \cite{hessels2006radio,nunez2010extracting,dubey2014analysis} and
 PSR B 1937+21 \cite{ashworth19831,nunez2010extracting,dubey2014analysis} for which the $r_{g}$ and `a' values
are (4.050km., 2.42325km.) and (4.050km., 2.19438km.) respectively. These two pulsars have been chosen arbitrarily, but we
tried our best to choose pulsars whose Schwarzschild radii($r_{g}$) and rotation parameters ($a$) values are as close as
that of Sun, so that the comparison will be easy. We take the impact parameters for them as the physical radii which are
20.1km.\cite{hessels2006radio} and 20.2km. \cite{ashworth19831} respectively. By using Eqn.(\ref{Eqn3cn845}) we find the values
of $\Delta \chi_{eq}$ to be 2\textdegree.2471 and 2\textdegree.0172, for the two pulsars PSR J 1748-2446ad and PSR B 1937+21
respectively. So the results are quite appreciable. These results have been listed in Table \ref{table1}. In Fig.\ref{Impact}, we plot $\Delta \chi_{eq}$  against $\chi_{eq}$  for the pulsar PSR  J 1748-2446ad  for various values of impact parameters from the centre of mass. The impact parameter values were taken as 1 time, 1.5 times and 2 times of the physical radius. We can clearly see the dependence of the rotation  of  polarization vector $\Delta \chi_{eq}$ n the initial value of polarization vector ( $\chi_{eq}$) . We observe this form Fig.\ref{Impact} rotation value attains a maximum, when initial position angle is 47 degrees approximately. From Eqn.(\ref{extreamumf})  also, we get the same  value of 47.25 degree  for the rotation value to be maximum. 
\begin{widetext}
\begin{table}[b]
\caption{\label{table1}
 Schwarzschild radius ($r_{g})$, Physical radius (R or b), Rotation parameter (a) and $\Delta \chi_{eq}$ as the total rotation of polarization vector
for different millisecond pulsars and Sun are shown.(original $\chi_{eq}=$45\textdegree, $\xi=1$ )}
\begin{ruledtabular}
\begin{tabular}{ccccc}
Name of Stars & $r_{g}$ (km.)& R(km.)& a (km.)& $\Delta \chi_{eq}$ \\
\hline
Sun & 3.000&0.7$\times 10^{6}$&1.69749& $4^{\prime\prime}.2873 \times10^{-6}$\\
PSR J 1748-2446ad \cite{hessels2006radio}&4.050&20.1&2.42325&2\textdegree.2471\\
PSR B 1937+21 \cite{ashworth19831}&4.050&20.2&2.19438&2\textdegree.0172\\
\end{tabular}
\end{ruledtabular}
\end{table}
\end{widetext}
As a further illustration of our calculation, we shall now consider prograde and retrograde orbits of light ray with respect
 to the rotation axis of the gravitational (or Kerr) mass. In case of prograde orbit the equation would
be same as Eqn.(\ref{Eqn3cn8}). But for the reverse or retrograde orbit, the rotational parameter (a)  would be taken as
 negative in Eqn.(\ref{Eqn3cn8}). Thus the expressions for rotation of polarization vector would be written as:

\begin{equation}\label{Eqn3cn9}
 \Delta \chi_{eq}(pro)= \arctan\frac{2\frac{r_{g}a}{b^2}}{2\frac{\sqrt{b}}{\sqrt{b-r_{g}}}-1}
\end{equation}
and
 \begin{equation}\label{3c2eq2}
 \Delta \chi_{eq} (retro)=- \arctan\frac{2\frac{r_{g}a}{b^2}}{2\frac{\sqrt{b}}{\sqrt{b-r_{g}}}-1}
 \end{equation}

We note that Eqn.(\ref{Eqn3cn9})is same as Eqn.(\ref{Eqn3cn845}) obtained earlier, for prograde. Calculated values for prograde and retrograde orbits differ only in sign or direction.
\subsubsection{Case III: $E_{z}=0$ i.e. $\xi= \infty$}
 Considering only the Z component is present between two electric field components ($\chi_{eq} $=90\textdegree). Now going back to Eqn.(\ref{Eqn3cn8})
we can show that $\lim_{\xi \rightarrow \infty}\Delta \chi_{eq}=0$ by applying L'Hospital rule. This is also evident from Fig.(2).
So in this case there will be no rotation of polarization vector as evident.
 At this stage, it will be worth to make a compassion with the work done by \cite{su1980effect} earlier. Those authors found that, for an asymptotic flat observer( GIF case), the rotation of polarization vector is solely due to the frame dragging effect. This is in perfect agreement with our observation in the present work. Further, \cite{su1980effect} considered distances beyond $6r_{g}$ and obtained the amount of rotation to be upto few degrees. This is also grossly in agreement with what we have observed. An exact comparison between the approaches and results obtained by \cite{su1980effect} with those by us, is not possible due to several reasons. The former authors, obtained numerical results by running some specific codes and sometimes { \it a very high degree of approximation} was applied. Also it appears that, they limited their rotation parameter values (in Kerr line element) lower than the Schwarzschild impact parameter. In addition, they obtained solutions under the cases where the impact parameter  (distance of closest approach of the light ray to the massive body) is at least six times the Schwarzschild radius. Further, the  authors also limited all their calculations, to an initial condition, where polarization of light ray is at 45 \textdegree with respect to the equatorial plane.\\
In our approach, all the above three restrictions are relaxed and the solutions that we obtained are of more general in nature. However, the underlying physics in the two approaches appear to be the same. This is obvious from the Equation (4.27) of  \cite{su1980effect} and our Eqn.(\ref{correq2}), where the rotation angle (of polarization vector) depends on the same set of metric elements and the mathematical nature of the dependence is the same.  In spite of the finer differences between the two approaches, we can make an attempt to compare the results obtained by the two methods as follows. As can be seen from Fig. (3) of \cite{su1980effect}, the rotation angle is about 2.5 degrees ( for GIF) under the conditions of  rotation parameter=  $r_g$  and impact parameter=6*$r_g$ .  With these same set of parameters, we can calculate from our Eqn.(\ref{Eqn3cn845}) the rotation of polarization vector to be about   2.69 degrees. Since our work, has fewer assumptions and approximations, we claim the results obtained by us to be more accurate. In addition, the present work is more general in nature, direct and has  more flexibilities to accommodate different geometries. 


\section{\label{level4}Polarization introduced in the light ray due to curved space-time}
In this section, we explore whether the effect of curved space-time introduces any polarization for light ray propagating through it or not?
Now from the definition of Stoke's parameter \cite{stokes1851composition,born1999principles}
for linearly polarized electromagnetic wave, one can write the expression for polarization (P) in terms of the Stokes vectors (Stokes\cite{stokes1851composition}) as:
\begin{subequations}
\begin{align}
 I & =<D_{y}^2>+<D_{z}^2> \label{4eq1a}\\
 Q & =<D_{y}^2>-<D_{z}^2> \label{4eq1b}\\
 U & =2<D_{y}D_{z}\cos{\delta}> \label{4eq1c}
 \end{align}
\end{subequations}
where $\delta$ is the phase difference between $D_{y}$ and $D_{z}$ and $<..>$ indicates the time average. In the present work,
 it is clear that that both the components $D_{y}$ and $D_{z}$ are in same phase. From Eqn.(\ref{correq1a})and Eqn.(\ref{correq1b})
 we can calculate $D_{y}$ and $D_{z}$, where it
has been assumed that the light is linearly polarized.
So from the above definition of polarization, we can write(omitting $<..>$ sign everywhere for our convenience),
\begin{equation}\label{2bbeq1}
 P=\frac{\sqrt{Q^{2}+U^{2}}}{I}=\frac{\sqrt{(D_{y}^{2}-D_{z}^2)^{2}+(2D_{y}D_{z})^{2}}}{D_{y}^{2}+D_{z}^2}=1
\end{equation}
Thus we find that, the light remains same way 100\% plane polarized, as it was before entering the gravitational
field. As a result, one can conclude here that the curved space time geometry (of Kerr field), does not introduce any new
polarization in the electromagnetic wave (or light). However, the gravitational field just rotates the direction of the
polarization vector (or electric vector).


\section{\label{result}Result and Discussions}

The above work clearly demonstrated and systematically calculated, the effect of curved space-time on the polarization states
 of a light ray. From this work, one can easily come to a conclusion that there is no effect on polarization, due to the gravitational field of static non-rotating body.
For example the Schwarzschild field will have no effect on the polarization of the
electromagnetic field so far, which is very obvious and supports the previous work by several authors. \\
Present work clearly indicates that, however the gravitational field generated by a rotating body
would affect the direction of polarization vector of electromagnetic wave.  The work presented here provides an analytical expression to find out the amount of rotation of polarization vector. This expression is much simpler and straightforward compared to the ones obtained by previous authors ; {\it viz} the work by \cite{su1980effect} requires many approximations and numerical techniques to find the exact value of rotation.  Calculations carried out here, for couple of
pulsars as test cases, demonstrated that the result is appreciable and it is in the order of few arc degrees, which can be
measured by present day telescopes. However, for Sun the calculated rotation is in the order of few micro-arcseconds, and this
can be measured by new generation telescopes and polarimeters. Also based on the calculations presented here, conclusions can
be drawn that the gravitational field would not produce or introduce any polarization in the light and it will only rotates the direction of
existing polarization vector.\\
 Our present study may also open new avenues for the indirect detection of gravitational waves (\cite{abbott2016observation}). This may be possible in two different ways (i) the electromagnetic counterpart of gravitational wave may be emitted in a way involving axial symmetry ({\it viz.} binary black hole coalescence \cite{abbott2016observation}), which will cause the rotation of polarization vector of electromagnetic wave passing close by; and/or (ii) the electromagnetic wave passing through a region of gravitational wave will be lensed \cite{sbytov1973collision}, resulting in a good possibility of rotation of its polarization vector, especially if the gravitational wave is plane polarized. However, these are only possibilities, which we would like to explore in our future work.

\section*{acknowledgments}

 Authors would like to thank Dr. B. Indrajit Sharma, Head of Department of Physics, Assam University, Silchar, India for useful discussions
 and suggestions. T. Ghosh is also thankful to Sanjib Deb, Anuj Kumar Dubey and Saumyadeep Roy Choudhury  Department
 of Physics, Assam University, for providing help and support in programming and making plots. The authors acknowledge the help from Inter University Center for Astronomy and Astrophysics (IUCAA), Pune, India for various helps it received towards this work. Finally, we are very much thankful to anonymous referees for their useful comments.
\bibliography{manuscript}

\providecommand{\noopsort}[1]{}\providecommand{\singleletter}[1]{#1}%
\begin{thebibliography}{}
\expandafter\ifx\csname natexlab\endcsname\relax\def\natexlab#1{#1}\fi

\bibitem[{Abbott {et~al.}(2016)Abbott, Abbott, Abbott, Abernathy, Acernese,
  Ackley, Adams, Adams, Addesso, Adhikari, {et~al.}}]{abbott2016observation}
Abbott, B., Abbott, R., Abbott, T., {et~al.} 2016, Physical review letters,
  116, 061102

\bibitem[{Ashworth {et~al.}(1983)Ashworth, Lyne, \& Smith}]{ashworth19831}
Ashworth, M., Lyne, A.~G., \& Smith, F.~G. 1983, Nature, 301, 313

\bibitem[{Balazs(1958)}]{PhysRev.110.236}
Balazs, N.~L. 1958, Phys. Rev., 110, 236

\bibitem[{Born \& Wolf(1999)}]{born1999principles}
Born, M., \& Wolf, E. 1999, Principles of optics: electromagnetic theory of
  propagation, interference and diffraction of light (Cambridge university
  press)

\bibitem[{Carroll(2004)}]{carroll2004spacetime}
Carroll, S.~M. 2004, Spacetime and geometry. An introduction to general
  relativity, Vol.~1

\bibitem[{Dubey \& Sen(2014)}]{dubey2014analysis}
Dubey, A.~K., \& Sen, A.~K. 2014, International Journal of Theoretical Physics,
  54, 2398

\bibitem[{Fern{\'a}ndez \& Davis(2011)}]{fernandez2011x}
Fern{\'a}ndez, R., \& Davis, S.~W. 2011, The Astrophysical Journal, 730, 131

\bibitem[{Frolov \& Shoom(2011)}]{frolov2011spinoptics}
Frolov, V.~P., \& Shoom, A.~A. 2011, Physical Review D, 84, 044026

\bibitem[{Hessels {et~al.}(2006)Hessels, Ransom, Stairs, Freire, Kaspi, \&
  Camilo}]{hessels2006radio}
Hessels, J.~W., Ransom, S.~M., Stairs, I.~H., {et~al.} 2006, Science, 311, 1901

\bibitem[{Jackson(1962)}]{jackson1962classical}
Jackson, J.~D. 1962, Classical electrodynamics, Vol.~3 (Wiley New York etc.)

\bibitem[{Kerr(1963)}]{kerr1963gravitational}
Kerr, R. 1963, Physical review letters, 237

\bibitem[{Landau \& Lifshitz(1998)}]{landau2013classical}
Landau, L.~D., \& Lifshitz, E.~M. 1998, The classical theory of fields, Vol.~2
  (Butterworth Heinemann)

\bibitem[{Misner {et~al.}(1973)Misner, Thorne, \&
  Wheeler}]{misner1973gravitation}
Misner, C.~W., Thorne, K.~S., \& Wheeler, J.~A. 1973, Gravitation (WH Freeman)

\bibitem[{Nunez \& Nowakowski(2010)}]{nunez2010extracting}
Nunez, P.~D., \& Nowakowski, M. 2010, Journal of Astrophysics and Astronomy,
  31, 105

\bibitem[{Paczynski(1986)}]{paczynski1986gravitational}
Paczynski, B. 1986, The Astrophysical Journal, 304, 1

\bibitem[{Pineault \& Roeder(1977)}]{pineault1977applications}
Pineault, S., \& Roeder, R. 1977, The Astrophysical Journal, 213, 548

\bibitem[{Plebanski(1960)}]{PhysRev.118.1396}
Plebanski, J. 1960, Phys. Rev., 118, 1396

\bibitem[{Roy \& Sen(2014)}]{Roy2014Sen}
Roy, S., \& Sen, A.~K. 2014, arXiv preprint arXiv:1408.3212v1

\bibitem[{Sbytov(1973)}]{sbytov1973collision}
Sbytov, Y.~G. 1973, Soviet Journal of Experimental and Theoretical Physics, 36,
  387

\bibitem[{Sen(2010)}]{sen2010more}
Sen, A.~K. 2010, Astrophysics, 53, 560

\bibitem[{Sereno(2004)}]{sereno2004gravitational}
Sereno, M. 2004, Phys. Rev. D, 69, 087501

\bibitem[{Stephani(2004)}]{stephani2004relativity}
Stephani, H. 2004, Relativity: An introduction to special and general
  relativity (Cambridge university press)

\bibitem[{Stokes(1852)}]{stokes1851composition}
Stokes, G.~G. 1852, Transactions of the Cambridge Philosophical Society, 9, 399

\bibitem[{Su \& Mallett(1980)}]{su1980effect}
Su, F., \& Mallett, R. 1980, The Astrophysical Journal, 238, 1111

\bibitem[{Vilenkin(1984)}]{vilenkin1984cosmic}
Vilenkin, A. 1984, The Astrophysical journal, 282, L51

\bibitem[{Visser(2007)}]{visser2007kerr}
Visser, M. 2007, arXiv preprint arXiv:0706.0622

\bibitem[{Weinberg(1972)}]{weinberg1972gravitation}
Weinberg, S. 1972, Gravitation and cosmology: Principle and applications of
  general theory of relativity (John Wiley and Sons, Inc., New York)

\end{thebibliography}



\listofchanges

\end{document}